\documentclass[10pt,a4paper]{article}
\usepackage{quantlaw_style}

\usepackage[numbers]{natbib}

\newcommand{\click}[2]{\href{http://#1}{\colorlet{temp}{.}\color{blue}{\underline{\color{temp}#2}}\color{temp}}}  
  
\usepackage[autostyle]{csquotes}

\begin{document}

\pagestyle{fancy}

\title{Graphie: A network-based visual interface for UK's Primary Legislation}
\author{Evan Tzanis}
\author{Pierpaolo Vivo}
\author{Yanik-Pascal F\"{o}rster}
\author{Luca Gamberi}
\author{Alessia Annibale}
\affil{Quantitative and Digital Law Lab, Department of Mathematics, King's College London,
	Strand, WC2R 2LS, London (United Kingdom) - \url{Quantlaw.co.uk}}

\maketitle
\thispagestyle{fancy}

\begin{abstract}
We present {\bf Graphie}, a novel navigational interface to visualize Acts and Bills included in the UK's legislation digital repository \textit{legislation.gov.uk}. {\bf Graphie} provides a network representation of the hierarchical structure of an Act of Parliament, which is typically organized in a tree-like fashion according to the content and information contained in each sub-branch. Nodes in {\bf Graphie} represent sections of an Act (or individual provisions), while links embody the hierarchical connections between them. The legal map provided by {\bf Graphie} is easily navigable by hovering on nodes, which are also color-coded and numbered to provide easily accessible information about the underlying content. The full textual content of each node is also available on a dedicated hyperlinked canvas. The building block of {\bf Graphie} is {\bf Sofia}, an offline data pipeline designed to support different data visualizations by parsing and modelling data provided by \textit{legislation.gov.uk} in open access form. While we focus on the Housing Act 2004 for illustrative purposes, our platform is scalable, versatile, and provides users with a unified toolbox to visualize and explore the UK legal corpus in a fast and user-friendly way. 
\end{abstract}

\section*{Keywords: legal data science, legislation, data pipelines, network interfaces, visualization of legal texts, user interface}

\clearpage
\pagestyle{fancy}
\section{Introduction}

The volume of the UK's primary legislation keeps growing at a very fast pace. According to a rough (and probably outdated) estimate, there are at least 176,890 \footnote{reported numbers till 2016, in \cite{LegislationVolume}} Public and General Acts currently in force in the UK - the exact number is not known - and an average of 30 new Public Acts are produced every year \cite{LegislationVolume}. New legislation \footnote{available here: \textit{\url{www.legislation.gov.uk/new}}} documents are regularly uploaded to the UK's Legislation web platform (\url{legislation.gov.uk}, \cite{UKLegislation}), managed by {\bf The National Archives} \footnote{\url{https://www.nationalarchives.gov.uk}} (TNA) on behalf of HM Government.

\begin{figure}
	\centering
	\includegraphics[width=0.8 \textwidth]{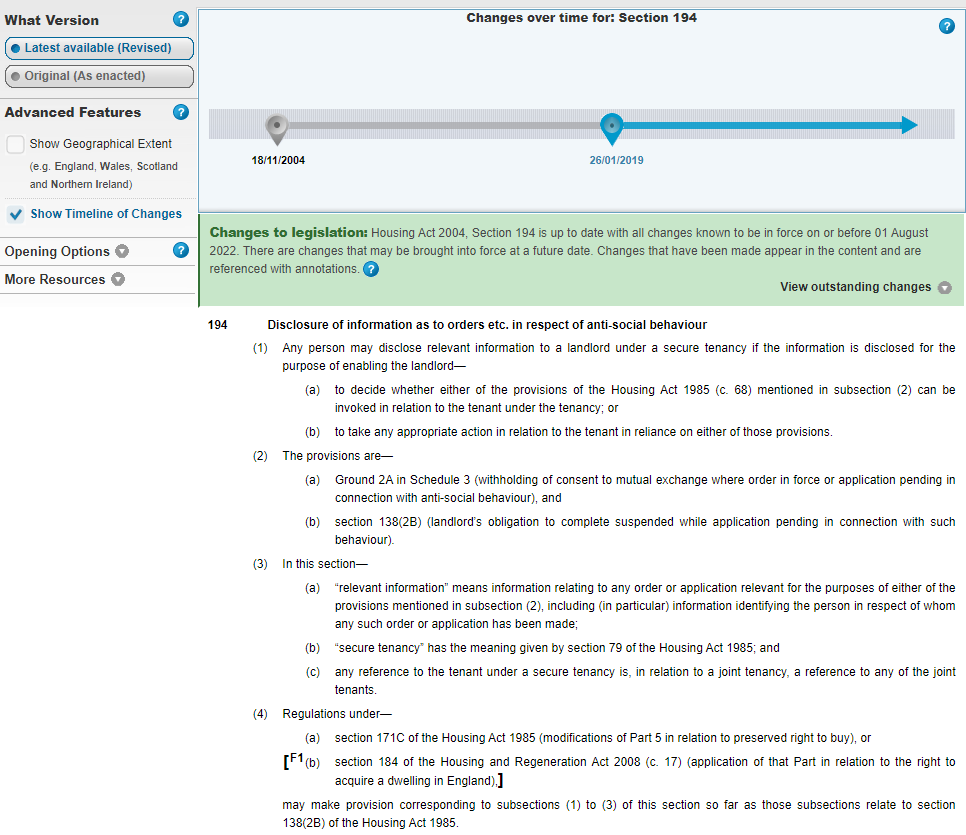}
	\caption{\label{fig:194} Section 194 of the Housing Act 2004 as provided by \cite{UKLegislation}. The default visualization offering includes (i) plain text of the provision (with limited hyperlinking) (ii) time-slider to access different versions of the provision (iii) choice between latest version, or version as originally enacted. }
\end{figure}

Users may reach the \url{legislation.gov.uk} webpage while looking for a specific Act or provision on standard search engines. Others may use the platform as part of their daily job. The \url{legislation.gov.uk} website has been carefully designed and is maintained to cater for the needs of a diverse pool of stakeholders. It is built on clear principles and offers a number of essential features: first, users can keyword-query the database, and are offered an easy-to-use set of navigational links for browsing through different corners of the UK legislation. Secondly, legislation data are open-source and fully accessible via an API \footnote{\label{footnote:api}the UK Legislation API, \url{https://www.legislation.gov.uk/index}}. All API legal documents are held in {\bf XML} format under a well defined and concise set of persistent URIs \footnote{\url{https://www.legislation.gov.uk/developer/uris}}. Thanks to this API technology and to TNA's open-access philosophy, the legislation data can also be connected and streamlined across other data sets and applications, such as for instance Westlaw \cite{Westlaw}, a leading commercial legal research platform. In addition, the \url{legislation.gov.uk} platform  enables users to enjoy the textual version of a whole Act -- or a section/paragraph thereof -- in HTML or in PDF formats. Acts are made available in both their original (as enacted) or revised (current) versions, and for those Acts with revisions, a detailed timeline highlighting any editing changes to legal documents over time is also provided (see fig.~\ref{fig:194}).

The standard set by TNA in terms of offering a digital and navigable version of essentially the entire corpus of UK legislation is very high and very competitive on the world stage. However, the lack of easy ``hopping'' capabilities between items and provisions that should be naturally linked together, as well as its focus on a text-only rendering of provisions leaves room for some improvement.

As for academic papers, reading and understanding legislation requires concentration and time, and the ability to efficiently `follow the leads' between different provisions of the same Act -- or between different Acts that have a bearing on the same matter. Consider again section 194 of the Housing Act 2004 as our main example, which is highly connected\footnote{This means that section 194 includes multiple references to other statutes.} with other sections from different Acts of the UK's Statute Book. To fully understand the content and implications of section 194, the reader is expected to visit and read the sections of these other statutes referenced there first, and then hop onto the sections/provisions that these other sections might refer to, and to repeat this hopping routine exhaustively, covering all possible linkages between sections/provisions/statutes. Using a text-based visualization interface with limited hyperlinking capabilities such as that provided by \cite{UKLegislation} makes these tasks time-consuming and inefficient for long and highly interconnected sections.

\begin{figure}
	\centering
	\includegraphics[width=0.99 \textwidth]{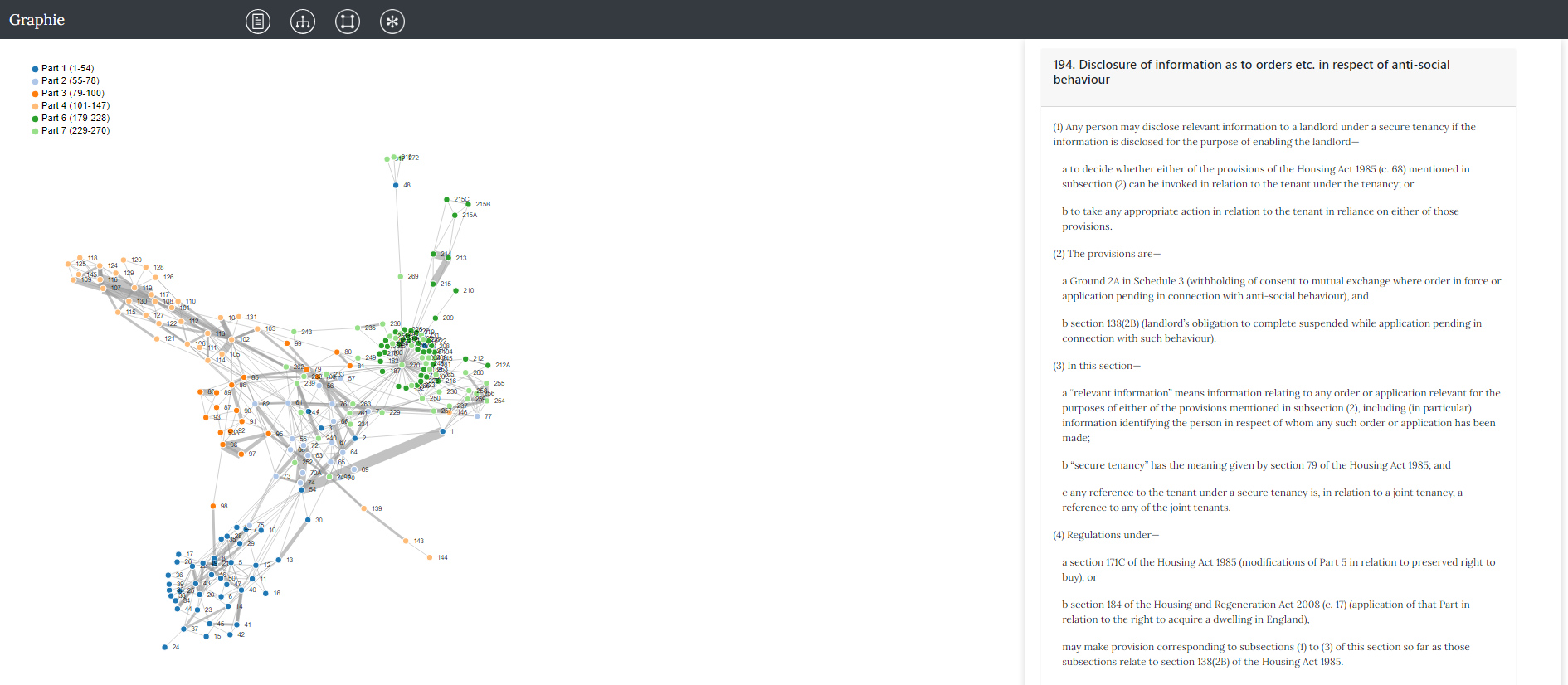}
	\caption{\label{fig:weight} An inbound weighted representation of the Housing Act 2004 in {\bf Graphie}. On the left, the web of provisions joined by a link whenever two of them can be reached in one-hop from one another. Nodes are color-coded (and the shape of the node marker can be changed too) to reflect -- in this particular case -- which Part of the Act each node belongs to. Hovering with the mouse over each node retrieves the textual content embedded in the node (on the right). Moving from each provision to its `neighbors' no longer requires refreshing/clicking on hyperlinks, but simply wandering around with the mouse over the nodes of interest.}
\end{figure}

Thus, there is a need for improved tools and visualizations to help both occasional and professional users manage potentially demanding explorations into legal documents. This is exactly the aim of {\bf Graphie} \footnote{\url{https://graphie.quantlaw.co.uk/}}, which provides a different and more attractive palette of network visualization tools (see an example in fig.~\ref{fig:weight}) that may prove useful for law researchers and practitioners, as well as for the general public.

Before describing the main technical features and capabilities of {\bf Graphie}, we put our enterprise in the wider context of Legal Map systems and similar initiatives, highlighting overlaps and differences.

\paragraph{Related Work} 
The philosophy and technical construction behind {\bf Graphie} do not write on a blank slate.
The concept of a {\bf Legal Map} system and its theoretical framework were already introduced in \cite{LegalMaps}. {\bf Legal Maps} are multi-layered systems offering an end-user experience similar to the user interfaces provided by geographic navigation systems, such as Google Maps. From a mathematical perspective, a {\bf Legal Map}  
is a directed multi-graph (network) that contains all the structural items of a given legal document \cite{LegislationNetwork}. {\bf Legal Map} visualizations could unlock dependencies between legal entities that may be difficult to extract otherwise, from identifying the most ``important'' nodes in a network, to clustering nodes according to a given notion of similarity. 

The use of tools from complexity science and network theory to analyze and represent legal texts has a relatively short but already fruitful history: arguably, the newly minted `Physics of the Law' field \cite{PhysicsoftheLaw} will play the same role to Law that ``Econophysics'' had to Economics \cite{EconophysicsReview}, and ``Biophysics'' to Biology (e.g. \cite{RandomWalksBiology}) in terms of cross-fertilization of ideas between distinct domains. An extensive graph-theoretic approach to the EU legislation network is given in \cite{LegislationNetwork}. In \cite{hierarchical} the tree-hierarchical network of the U.S. Code is examined by considering several scoring and ranking metrics. Ref. \cite{LegalTrees} builds a hierarchical model of information (distributed on the nodes of a tree) and defines a notion of ``structural complexity" on the basis of the average time a random reader takes to retrieve some piece of information planted in the leaves. 

The authors in \cite{LawTime}, prior to applying a network-driven analysis against the statutes and regulations in the United States and Germany, perform several pre-processing steps on their underlying raw data. Legal documents are usually found as texts with little metadata, which makes them hard to use in applications of legal informatics \cite{LegalInformatics}. We undertake a similar data preparation exercise against our raw data in section \ref{sec:datapipeline}. 

Network-based representations of ``information'' do exist in other contexts, for instance academic papers. Scholarly archiving systems (Semantic Scholar \footnote{\url{https://www.semanticscholar.org/}}, PubMed \footnote{\url{https://pubmed.ncbi.nlm.nih.gov/}}, Arxiv  \footnote{\url{https://arxiv.org/}}) use state-of-the-art AI and API engineering that make scientific documents' processing easier. This enables the development of applications such as Connected Papers \footnote{\url{https://www.connectedpapers.com/}}. Connected Papers is a network-driven citation analysis tool for enabling end users to explore relevant academic papers. The tool facilitates collecting and analyzing academic references from the chosen archiving system. Network visualizations in Connected Papers are developed using D3.js \footnote{\url{https://d3js.org}}, a well-established JavaScript library for producing bespoke and interactive visualizations. With {\bf Graphie}, we aim to develop a similar tool, using the same front end technology, but tailored to a different -- and arguably less malleable -- type of raw data. Indeed, the XML versions of the legal texts provided by \cite{UKLegislation}'s API have an intricate and complex sub-structure, which is markedly different from the short-text nature (say, titles and abstract) typical of academic papers handled by citation APIs. Thus, {\bf Graphie} faces the extra challenge of having to parse and build a unified model of long and intricate legal documents starting from their (XML) raw representation. 

Coming back to the legal platforms field, LAWSampo \cite{LawSampo} is an example of a modern legal semantic web portal in the context of the Finnish Legislation. LAWSampo is built according to the FAIR \cite{Fair} principles of the Sampo Model \cite{SampoModel} and the SAMPO-UI \cite{SampoUI}, a full-stack Javascript framework. LAWSampo's architecture clearly separates the user interface (SAMPO-UI) from the underlying Linked Data service via a SPARQL API \footnote{\url{https://www.w3.org/TR/sparql11-query/}}. Software developers or legal analysts could use SPARQL endpoints and query LAWSampo's data service for their own Python or R applications (say network visualizations), using Jupyter notebooks. SAMPO-UI supports network visualizations by including a Cytoscape.js \footnote{\url{https://js.cytoscape.org/} - A Graph theory library} based component, already demonstrated in few portal instances of the Sampo model, existing in LetterSampo \footnote{ \url{https://lettersampo.demo.seco.cs.aalto.fi/en/actors/faceted-search/network} LetterSampo, A network of Historical Letters}  and in AcademySampo \footnote{\url{https://akatemiasampo.fi/en/people/faceted-search/network} AcademySampo, A network of Finnish Academic People}. In {\bf Graphie}, we aim to develop the ``Visualization'' feature, also mentioned in~\cite{GovernmentData}, where sections/provisions of one Act are represented as nodes, and their ``connections'' along the information hierarchy as edges. Each node is endowed with its own primary XML schema reference (described below). 

To facilitate the web development of LAWSampo, the authors in \cite{LawSampo} completed a specific data exercise by initially transforming legal documents hosted on Finlex's Data Bank server Finlex into a Linked Open Data (LOD) repository, named Semantic Finlex \cite{SemanticFinlex}. Consequently, Semantic Finlex's data were converted into a data format compatible with LAWSampo's semantic portal. While we do not use the concept of LOD in our work, we took a similar pre-visualization data processing exercise in {\bf Graphie}. Prior to any visualization, in section \ref{sec:datapipeline} we illustrate how XML raw data from \cite{UKLegislation} are parsed and then checked against data quality indicators. The Graphie Data Model is not strictly following the FAIR principles, as it is not published as an open ontology. 

\section{Sofia: A cross Act pipeline} \label{sec:datapipeline}

\begin{figure}%
    \centering
    \includegraphics[width=0.99 \textwidth]{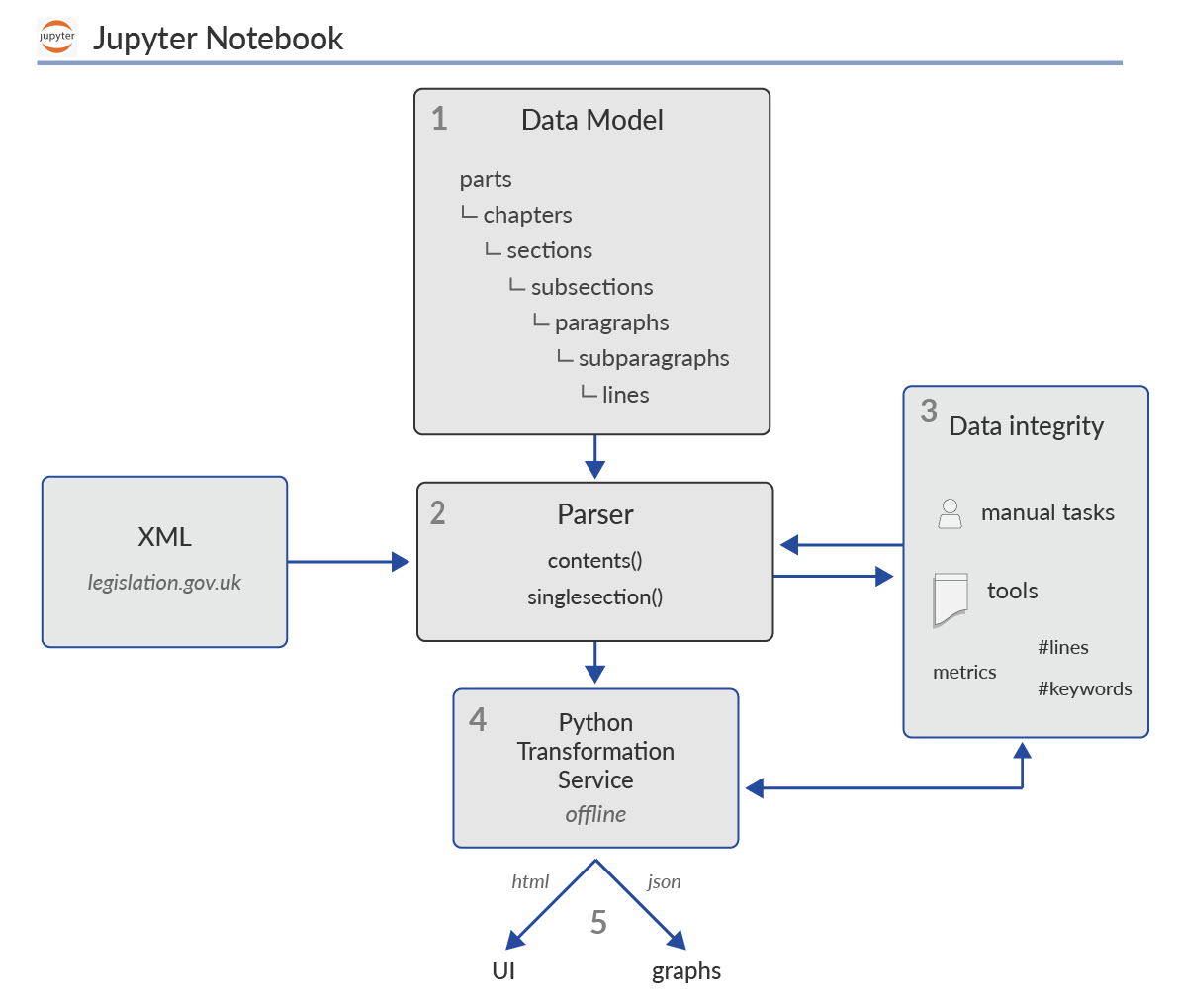}
    \caption{Sofia is an offline, Python-built pipeline developed as a Jupyter notebook. To implement this pipeline, we apply a multi-phase procedure that includes the following: Data Modelling, Parser, Data Integrity, Transformation Service and Visualizations. Data Integrity routines are performed and hosted outside the mentioned Jupyter page.}%
    \label{fig:pipeline}%
\end{figure}

{\bf Graphie}'s aim is to tame the complexity of long and intricate legal texts by departing from the traditional ``text$+$hyperlinks'' philosophy adopted by most digital archives, in favor of a more holistic, network-based representation of the underlying information content. We achieve this goal by defining the following multi-phase pipeline (see figure \ref{fig:pipeline}):

\begin{enumerate}
    \item  We codify and represent one Act's hierarchical structure and its textual content using a programming language (in our case, Python). In section \ref{sec:graphiedatamodel}, a Python object is accordingly declared, henceforth named: Graphie Object (or Graphie's Data model). (\textit{\textbf{Data modelling}})
    \item  We parse the raw XML document of one specific Act using the XML parser defined in section \ref{sec:parser}. Ingested data feed an instance of the Graphie Object, defined in phase 1. (\textit{\textbf{Parser}})
    \item  We undertake several data integrity procedures against the parsed textual elements of the Graphie object from phase 2 to ensure their quality and completeness. (\textit{\textbf{Data Integrity}})
    \item  We convert, using Python, the obtained instance of the Graphie Object from phase 3 into specific JSON files or HTML components. (\textit{\textbf{Transformation Service}})
    \item We use the JSON files and the HTML components from phase 4 for finally feeding the underlying network libraries and certain HTML parts of our platform. (\textit{\textbf{Visualizations}})
\end{enumerate}

We designed this pipeline with data analysts in mind. Thus, Sofia is implemented as a Jupyter notebook, which offers a streamlined experience for capturing and processing UK Legislation XML documents. As we discuss in section \ref{sec:dataintegirty}, data checking tools and activities are hosted outside the aforementioned Jupyter notebook and based on the outcome, the XML parser component of phase 2 is expected to be adjusted accordingly. For this reason, Sofia is an example of an offline pipeline in which we carry out data integrity checks on outputs of phase 2 and phase 4. In the following subsections we describe in detail the different phases of our pipeline, before discussing future work in  section \ref{sec:conclusions}. 

\subsection{Data Modelling} \label{sec:modelling}

Prior to any graph visualization, data modelling of the original data is required. In our case, we need to map elements of the UK Legislation XML files into a target data model, which is closer to our final network representation. This requires identifying: which elements from the original XML dataset should act as nodes, under which relation two nodes should be linked to one another, and whether we display any other numerical or descriptive information over our graph's edges and nodes by applying related visual effects. The following paragraphs outline raw data's entities, relationships and properties, which we codify as the {\bf Graphie}'s data model in Python. 

\paragraph*{Primary Data}

\begin{figure}%
    \centering
    \includegraphics[width=0.99 \textwidth]{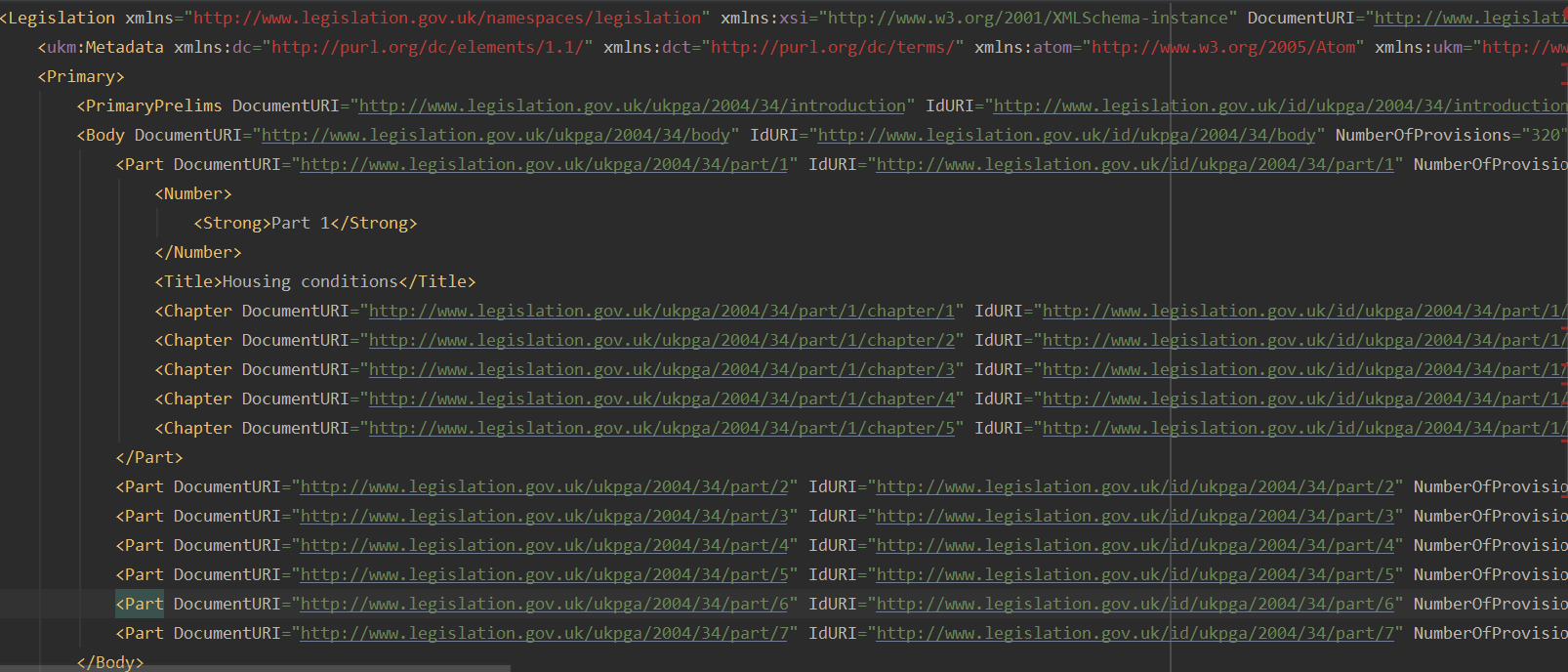}
    \caption{ The XML tree representation of the Housing Act 2004 \cite{HousingActXML} contains 7 tags named \texttt{Parts} (corresponding to the 7 parts of the same Act), and each tag \texttt{Part} embeds other tags, to name a few : \texttt{Chapters, Pblock, Title, P1group}. Subsequent \texttt{P1group} tags indicate the start of a section (see fig. \ref{fig:xmlsection})}%
    \label{fig:parts}%
\end{figure}

\begin{figure}%
    \centering
    \includegraphics[width=0.99 \textwidth]{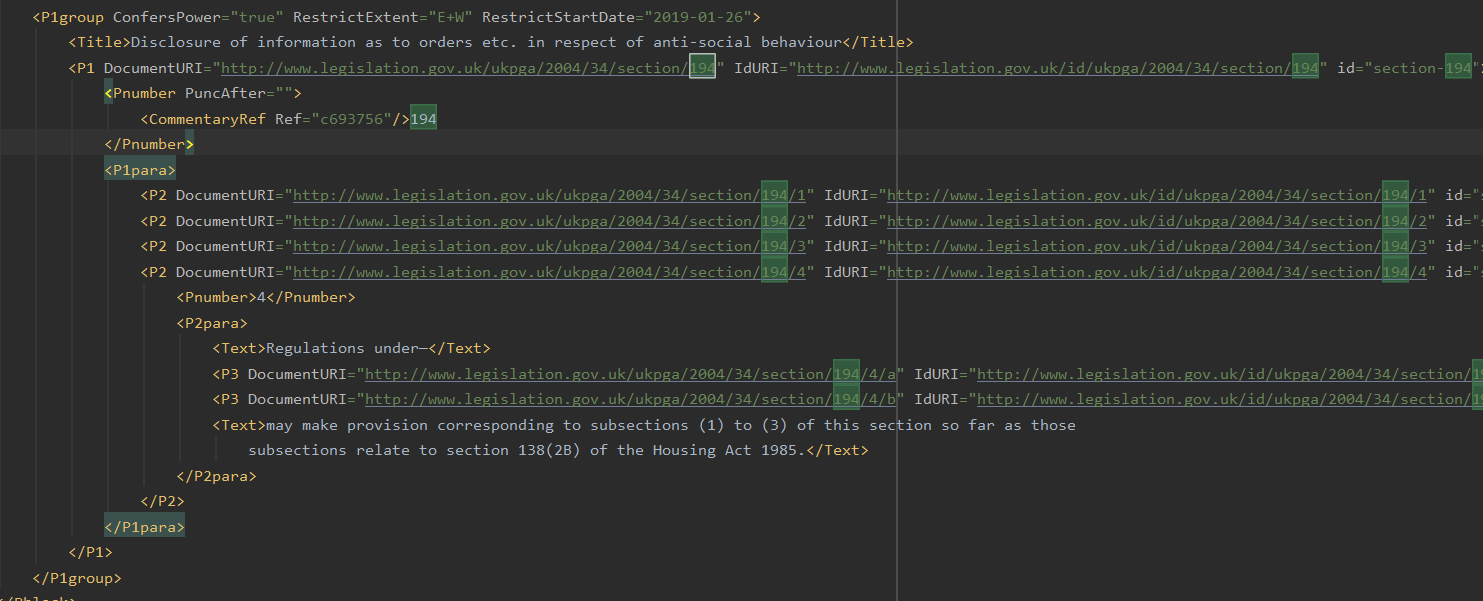}
    \caption{ Section 194 of the Housing Act 2004, within the XML file \cite{HousingActXML}. The content of this section is formed in a sequence (section specific) of embedded tags: \texttt{P1group, P1, P1para, P2, P2para, P3, Text}. Other sections, say section 193, due to their textual complexity, might have a different tag ordering, involving \texttt{BlockAmendment} tags. }%
    \label{fig:xmlsection}%
\end{figure}

Each legislation page (either a whole item, or a part, or a section) in \cite{UKLegislation}, is also offered as an XML file, which we refer to as the XML URL of that page. For instance, the full data of the Housing Act 2004, hosted at \url{https://www.legislation.gov.uk/ukpga/2004/34/}, is also available at \url{https://www.legislation.gov.uk/ukpga/2004/34/data.xml} \footnote{Which we call the full data XML file, also used in section 3.2.1}. {\bf Legislation XML} files use Crown Legislation Markup Language's (CLML) syntax and the associated schema. The XML files of the {\bf UK Legislation API} are structured in two essential layers, the metadata layer and the content layer. Quoting from \cite{CLML}, ``the CLML model incorporates versioning and facilitates the notion of expressing changes over time which helps us to understand the underlying metadata semantics behind the surface content''. The metadata layer is formed by elements such as: \texttt{title, publication, type, format} and a rigid set of persistent URIs. The content layer contains a mix of {\bf hierarchical} (\texttt{Parts, Chapters}) (see fig. \ref{fig:parts}) and {\bf non-hierarchical, textual} tags (\texttt{Pblock, P1group}) (see fig. \ref{fig:xmlsection}). In this paper, we only focus on the content layer. {\bf CLML} schema's tag complexity is explored in \cite{tamingthecomplexity} and also shown in Fig 3, on page 13 in \cite{CLML}. 

In {\bf Graphie} we wish to visualize Acts as networks, where each section is connected with other sections if and only if there is clear textual reference between them. Consider the following last line from section 194 ``... so of this section so far as those subsections relate to section 138(2B) of the Housing Act 1985...''', in figure \ref{fig:networkgraphs}. In the same figure, the mentioned dependence is pictured as a link connecting the node representing the section 194 and the node representing the Housing Act 1985. 

In the following section we define a minimal data model, sufficient for capturing both the {\bf hierarchical} and {\bf non-hierarchical, textual} elements of one Act's XML structure.

\paragraph*{Graphie Data Model (or Graphie Object)}  \label{sec:graphiedatamodel}

The UK Primary legislation is structured in Acts and each Act includes different levels of division (such as Parts, Chapters, cross-headings) in a certain order, a hierarchy outlined in \cite{LegislationStructure}. The Housing Act 2004 in figure \ref{fig:contents} includes seven numbered parts. Each Part contains numbered Chapters and cross-headings. Chapters are above cross-headings. Cross-headings are not numbered and are displayed in italics and highlight a group of sections beneath it. Part 1, as an example, contains Chapters, Chapters include cross-headings and each cross-heading contains sections. Part 3 instead, contains cross-headings but not Chapters. 

Within the structural hierarchy defined in \cite{LegislationStructure}, sections are the lowest level of a piece of legislation. On the textual level, they contain sub-items organized in paragraphs that might include further numbered items. This textual structure is well captured by the related definitions from section 16 (page 21) in \cite{ReadingLegislation}. Section 194 (figure \ref{fig:194}) consists of: a number (194), a heading (``Disclosure of information as to orders...'') and its content. Section 194's content is divided into four numbered (1), (2), etc subsections. Each subsection starts with an introductory line and then contains two or more paragraphs. Subsection (3) contains three letter-ordered paragraphs. Subparagraphs are usually numbered `(i), (ii)...' and are nested within Paragraphs. For instance, consider section 97 \footnote{\url{www.legislation.gov.uk/ukpga/2004/34/section/97}} and its subsection (6) that contains two subparagraphs (i), (ii) within paragraph (b). Conjunctive words, such as ``and'' or ``or'', might be used for transitioning between paragraphs and subparagraphs. The canonical fine substructure of a section is suggested in \cite{ReadingLegislation}; such a refined definition is however missing from \cite{LegislationStructure}. The UK's Legislation XML documents, with few exceptions, are structured according to \cite{ReadingLegislation}'s section-text logic. 

In figure \ref{fig:pipeline}, the legal hierarchy of \cite{LegislationStructure} is included within the Data Model component and is represented by the following objects: \texttt{Parts}, \texttt{Chapters} and \texttt{cross-headings}, whereas the textual structure mentioned in \cite{ReadingLegislation} is embedded within the object \texttt{Sections} and includes a four-level object hierarchy:  \texttt{SubSections}, \texttt{Paragraphs}, \texttt{SubParagraphs} and \texttt{Lines}. Lines are the lowest element of this textual hierarchy. Thus, instances of the {\bf Graphie} model meet the three specific graph model criteria (\textit{hierarchy}, \textit{sequence}, \textit{reference}) outlined in \cite{LawTime} (page 2): 

\begin{enumerate}
    \item   Elements included in one instance of the {\bf Graphie} model are hierarchically structured. 
    \item   One element's text value is always embedded within an object of a higher-level position. These objects should also be sequentially ordered. 
    \item   One element's text might contain words for expressing cross-references to other sections of the same Act ({\bf inbound references}) or for pointing to sections from other Acts ({\bf outbound references}). 
\end{enumerate}

We implement the details of the third criterion in section \ref{sec:transformation}. Methods \texttt{refInSection()} and \texttt{actsInSection()} are used for capturing one section's cross-references and citations, respectively.

\subsection{Parser} \label{sec:parser}

In this section, we will walk through the main methods of the Python XML parser given in figure \ref{fig:pipeline}.
The parser is developed using the \texttt{beautifulsoup} \footnote{\url{https://www.crummy.com/software/BeautifulSoup/}} library, a well known Python tool, which allows one to try out different web scrapping strategies. 

\begin{figure}
	\centering
	\includegraphics[width=0.8 \textwidth]{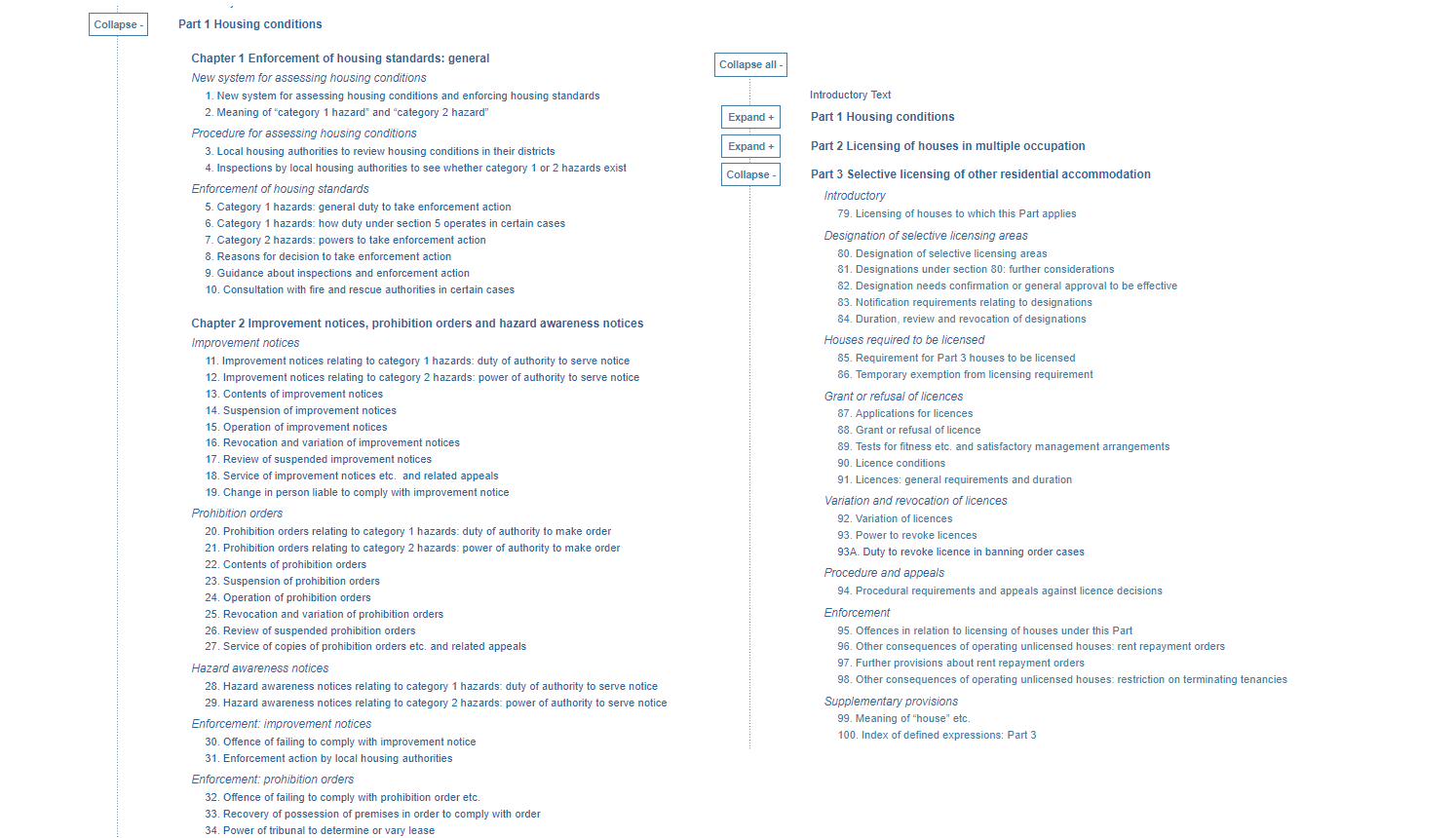}
	\caption{\label{fig:contents} Housing Act's 2004 table of contents as provided by \cite{UKLegislation}. The Act is organized in 7 Parts, following the Legislation Structure rules of \cite{LegislationStructure}. Sections in Part 1, shown on the left side, are located below cross-headings. Chapters in Part 1 include a series of cross-headings. Part 3 contains only cross-headings, but no Chapters. }
\end{figure}

Consider the ``table of contents'' page (figure \ref{fig:contents}) of the Housing Act 2004 here: \url{https://www.legislation.gov.uk/ukpga/2004/34/contents}. As expected, both the underlying XML document and the displayed HTML page are structured in Parts, Chapters and cross-headings. Each section's titles are hyperlinked, and each hyperlink points to one section's whole web document. For instance, in figure \ref{fig:contents}, section 3's title ``Local housing authorities to review housing conditions in their districts'' is located below the cross-heading ``Procedure for assessing housing conditions'' and points to the following page: \url{https://www.legislation.gov.uk/ukpga/2004/34/section/3}, the individual web page for section 3. 

The two fundamental units of our parser are: the \texttt{contents()} method (section \ref{sec:contents}) and the  \texttt{singleSection()} method (section \ref{sec:singlesection}). Each method can be executed in isolation. The role of \texttt{contents()} is to map the legal hierarchy structure included in one Act's ``table of contents'' XML page into the following items of the Graphie Object: \texttt{Parts, Chapters, cross-headings} (mentioned in section \ref{sec:graphiedatamodel}). Given one section's XML URL and using the \texttt{singleSection()} method instead, we can capture and map the textual content of one section to the following elements of the Graphie Object: \texttt{Sections, Paragraphs, SubParagraphs} and \texttt{Lines}. Both methods are not Act-specific and will be discussed elsewhere in this paper as well, such as in the context of Data Integrity (section \ref{sec:dataintegirty}) and in the Transformation Service (section \ref{sec:transformation}) sections. 

\subsubsection{contents()} \label{sec:contents}

\begin{figure}
	\centering
	\includegraphics[width=0.8 \textwidth]{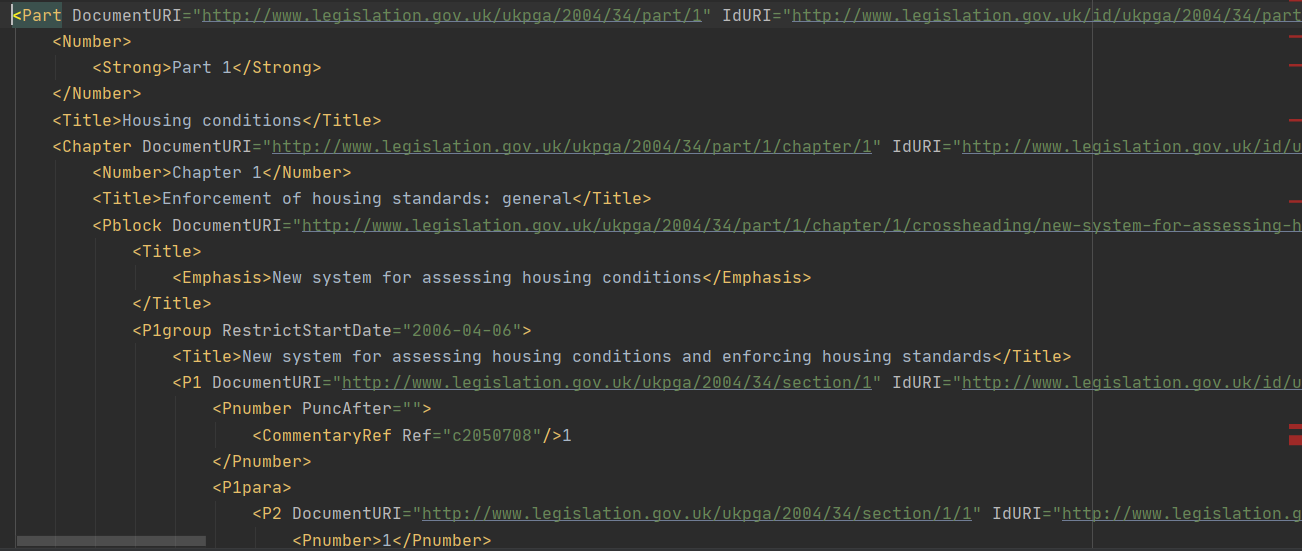}
	\caption{\label{fig:xmlconents} Housing Act 2004 - the XML representation of the full data XML file as provided by \cite{UKLegislation}. The file contains information about same Act's components: Parts, Chapters, cross-headings, and sections.}
\end{figure}

The ``Table of Contents'' page is where legislation readers could get an idea of how an Act is organized according to the legal hierarchy outlined in \cite{LegislationStructure}. The ``Table of Contents'' web page for the Housing Act 2004 \footnote{\label{footnote:housing_act_2004_contents}\textit{\url{https://www.legislation.gov.uk/ukpga/2004/34/contents}}} is divided into Parts, Chapters or cross-headings, and sections. All sections have headings alongside their numbers, and are grouped within a Chapter or a cross-heading. The above structure is well identified in figure \ref{fig:xmlconents}, corresponding to one Act's full data XML file: A \texttt{Part} is the root element for a group of \texttt{Chapters}. One \texttt{Part} element has a number and a title. \texttt{Chapter} tags have numbers and titles, too. A \texttt{Chapter} often includes cross-headings. A \texttt{Pblock} start tag indicates a cross-heading, and each cross-heading has a name. A cross-heading usually contains two or more sections. Sections are defined as \texttt{P1group} tags and always contain other elements such as \texttt{Title}, \texttt{PNumber}  and \texttt{DocumentURL} elements. 

\begin{figure}%
	\centering
	\includegraphics[width=1 \textwidth]{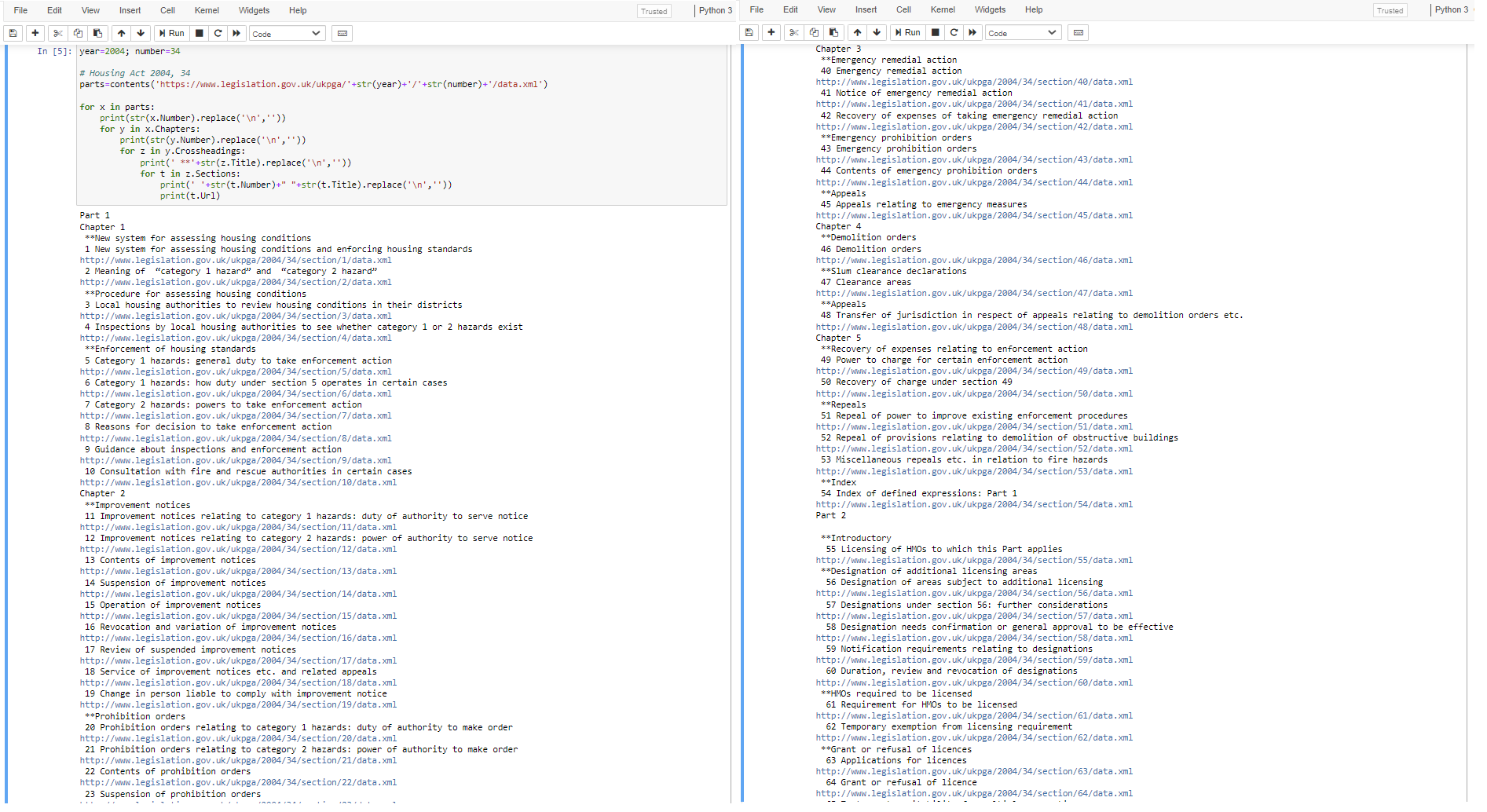}
    \caption{Housing Act 2004 - Table of Contents. {\bf contents()} is a function with one argument \texttt{url}.  When {\bf contents()} is called, we pass along the URL corresponding to one Act's full data XML file.
    This URL is used inside the function for returning an instance of the Python object \texttt{Parts} that contains nested occurrences of the following objects: \texttt{Chapters}, \texttt{cross-headings} and \texttt{Sections}. 
}%
    \label{fig:parsing}%
\end{figure}

For indexing one Act's legal structure, we use the method \texttt{contents()}. The parsed table of contents of the {\bf Housing Act 2004} is shown in figure ~\ref{fig:parsing}. The method \texttt{contents()} is clearly able to handle the structural variety between Part 1 and Part 2 (no chapters, only cross-headings). 

At this stage, one can easily declare a new array variable, say \texttt{urls}, for storing all the \texttt{DocumentURL} elements returned by \texttt{contents()} (see figure~\ref{fig:parsing}). In the next section we discuss how we could fully XML-scrap one Act's content by running the next defined method \texttt{singleSection()} on each \texttt{DocumentURL} in \texttt{urls}. 

\subsubsection{singleSection()} \label{sec:singlesection}

\begin{figure}%
    \centering
    \includegraphics[width=0.95 \textwidth]{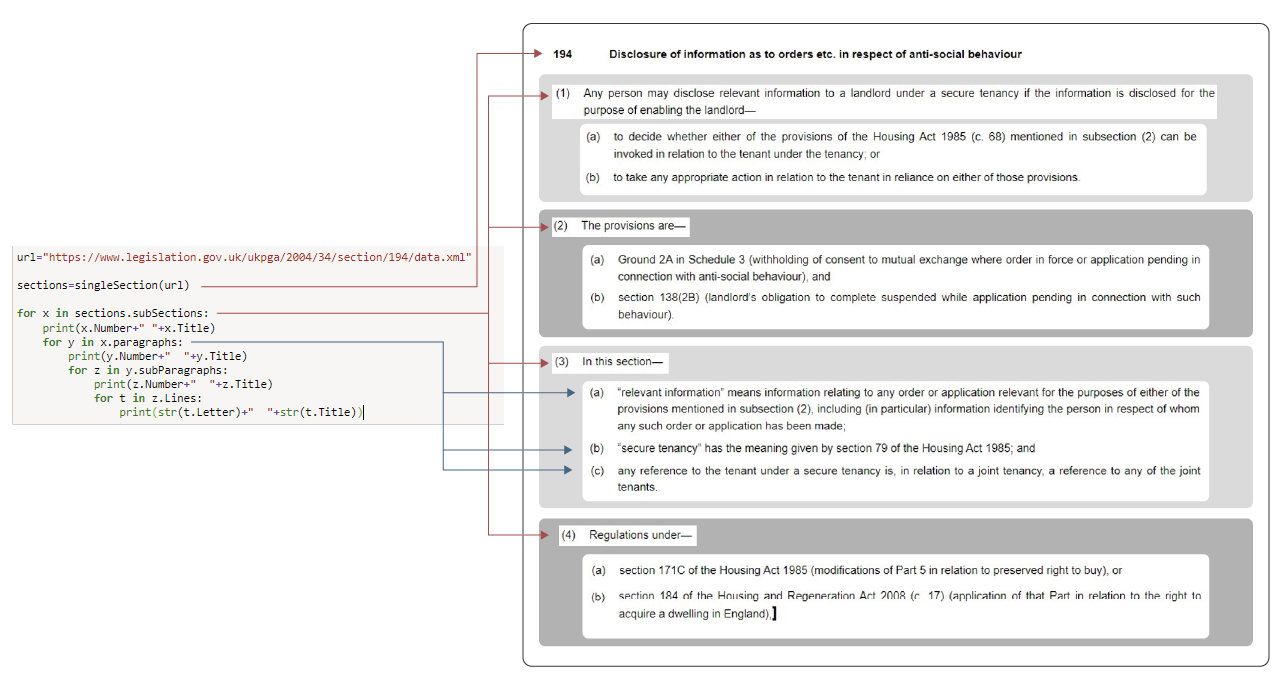}
    \caption{The textual on-memory representation of section 194 after parsing its content using the method \texttt{singleSection()}. The obtained \texttt{Sections} instance contains embedded objects (\texttt{SubSections}, \texttt{Paragraphs}, \texttt{SubParagraphs} and \texttt{Lines}) which are hierarchical and sequentially ordered: section 194 contains 4 subsections. Each subsection includes ordered paragraphs.}%
    \label{fig:singlesection}%
\end{figure}

The aim of the \texttt{singleSection()} method is to map, applying the text-structural rules of \cite{ReadingLegislation}, one section's (or subsection's) content into an instance of the Python object \texttt{Section} defined in section \ref{sec:graphiedatamodel}. In figure~\ref{fig:singlesection}, the object \texttt{Section} captures section's 194 content in four instances of the object \texttt{SubSections}, reflecting the four numbered items in figure \ref{fig:194}. Each instance of \texttt{SubSections} consists of an array of \texttt{Paragraphs} objects, corresponding to the letter-ordered lines. In our example, the \texttt{SubSections} object of the third item is composed of three \texttt{Paragraphs}, each paragraph corresponding to the ordered list (a), (b) and (c). 

Following legislation amendments, one section's subsection can modify another Act's sections content \cite{ReadingLegislation}. Consider that due to the use of the word ``insert'' in subsection (4) of section 185\footnote{\url{www.legislation.gov.uk/ukpga/2004/34/section/185}} of the Housing Act 2004,  section 155 of the House Acting Act 1985 is modified by inserting two new sections, sections 155A and section 155B. The word ``substitute'' (see subsection (2) in section 185) is also used for such amendments. In our data model (see figure \ref{fig:pipeline}) sections embed subsections and subsections incorporate paragraphs. Since amendments modify sections, they may be seen as superior objects to sections; on the other hand, their text is usually located within subsections. Our data model does not allow for sections to be subordinate to subsections, and is therefore unable to fully express amendments (see section \ref{sec:dataintegirty}). 

{\bf singleSection()} is heuristic in nature and depends on the underlying structural variety of the {\bf Legislation XML} schema. For achieving high levels of accuracy, {\bf singleSection()}'s output should be manually or programmatically checked, as we explain in the next subsection. 

\subsection{Data Integrity} \label{sec:dataintegirty}

The aim of the XML parser developed in section \ref{sec:parser} is at least twofold: firstly, we wish to speed up the data collection process and secondly, we plan to use the same parser for collecting data from other Acts. Data integrity is captured as a dedicated component in the flowchart of figure \ref{fig:pipeline}, a component that requires data analysts well-versed in the use of various tools. In our scenario, prior to any visualization, we are looking to improve our XML parser's output by incorporating related data quality observations. 

\begin{figure}%
    \centering
    \includegraphics[width=0.6 \textwidth]{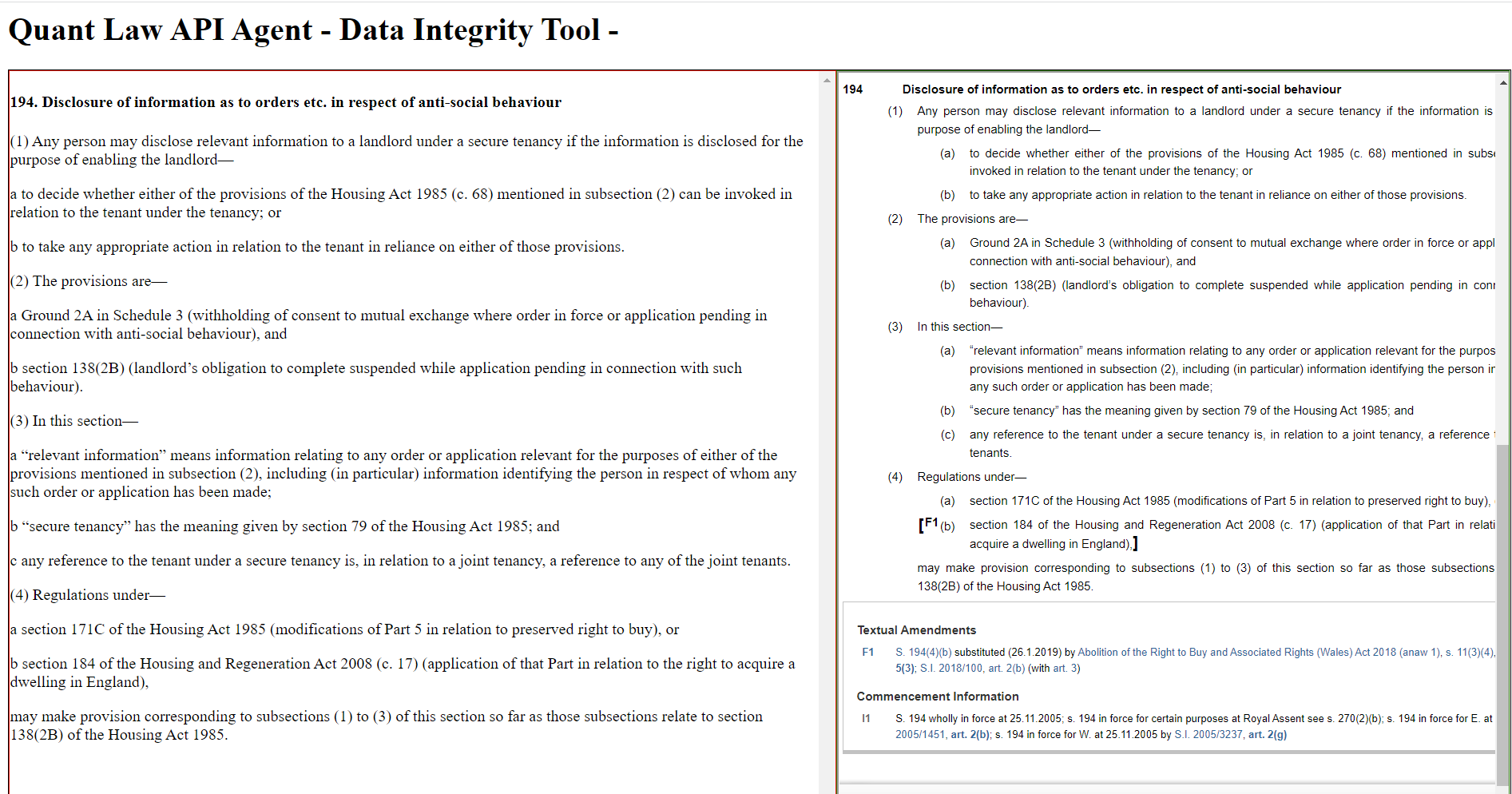}
    \caption{ A screenshot from the bespoke data quality tool, developed for improving quality checks based on visual comparison on two different monitors. On the left, section 194's parsed content. On the right, same section's content in \cite{UKLegislation}. }%
    \label{fig:dataintegrity}%
\end{figure}

Consider a data quality engineer parsing the XML file of section 194 of the Housing Act 2004. Following the steps of our pipeline, they would need to inspect {\bf singlesection()}'s output (displayed on a Jupyter window) against the original content of the same section on \url{legislation.gov.uk} (an HTML web page). Practically, this would require a visual comparison between two different pages on a browser. For facilitating speedy comparisons between parsed and original data, we developed a bespoke data integrity web tool. In figure \ref{fig:dataintegrity}, the left frame displays one section's parsed content. The same section's original content (shown in \cite{UKLegislation}) is located on the right frame. This blended visualization between original and parsed data makes their comparison easier and faster. Otherwise, we would have to keep two full screens open for comparing parsed and original data. Other key variables that we could use for evaluating our parser's performance are the total number of subsections, paragraphs, sub-paragraphs and single lines included within one section. 

Regular expression tools are also very useful for data engineers.  In table \ref{table:dataintegritytable} we report the frequency of a few important expressions after running Regex Search \footnote{\label{lognote:regexsearch} \textit{\url{https://chrome.google.com/webstore/detail/chrome-regex-search/bpelaihoicobbkgmhcbikncnpacdbknn}}} on the Housing Act's 2004 one-page view \footnote{\label{footnote:housing_act_2004}\textit{\url{https://www.legislation.gov.uk/ukpga/2004/34}}}. Consider the regular expression ``\textit{of the [aA-zZ]+ Act}". This regex is used for finding all Act names mentioned within the Housing Act 2004 and reports 156 such instances (outbound references). In table \ref{table:externalacts}, prior to the network visualization in figure \ref{fig:networkgraphs}, we match these instances with the output obtained by our parser's method: \texttt{actsInSection()} (see section \ref{sec:transformation}). ``\textit{section [1-9]* or [1-9]}" identifies those lines (say, subsection 1 in section 13 \footnote{\url{www.legislation.gov.uk/ukpga/2004/34/section/13}}) which include a section number (inbound reference), an `or' and are followed by a numerical digit from 1 to 9, pointing to other sections. For the network shown in figure \ref{fig:weight}, this is a useful pattern as it helps us to pin down and review edges corresponding to sections related by an `or' conjunction.

\begin{table}[ht]
\centering
    \begin{tabular}{ | l | l | p{5cm} |}
    \hline
    {\bf Regex}  & {\bf Occurrences} & {\bf examples}  \\ \hline
    \textit{section [1-9]$^{*}$} & 500 &  \click{www.legislation.gov.uk/ukpga/2004/34/section/1}{1(2b)} \\ \hline
    \textit{section [1-9]*} or [1-9] & 20 &  \click{www.legislation.gov.uk/ukpga/2004/34/section/13}{13(1)} \\ \hline
    \textit{section [1-9]* or section [1-9]} & 3 &  \click{www.legislation.gov.uk/ukpga/2004/34/section/102}{102(10)} \\ \hline
    \textit{section [1-9]* to} &  21 &  \click{www.legislation.gov.uk/ukpga/2004/34/section/1}{1(5)} \\ \hline
    \textit{sections [1-9]*} and & 4  &  \click{www.legislation.gov.uk/ukpga/2004/34/section/105}{105(11)} \\ \hline
     \textit{of the [aA-zZ]+ Act} & 156 & \click{www.legislation.gov.uk/ukpga/2004/34/section/1}{1(5)} \\ \hline
    \end{tabular}
    \caption{\label{table:dataintegritytable} Results of single line regular expression searches, executed using Regex Search (see footnote \ref{lognote:regexsearch}) on one page's view of the Housing Act 2004. A data engineer can use these patterns as alternative metrics for reviewing our parser's performance about cross-referencing and citations collection.}
\end{table}

\subsection{{\bf Transformation Service}} \label{sec:transformation}

\begin{figure}
	\centering
	\includegraphics[width=1 \textwidth]{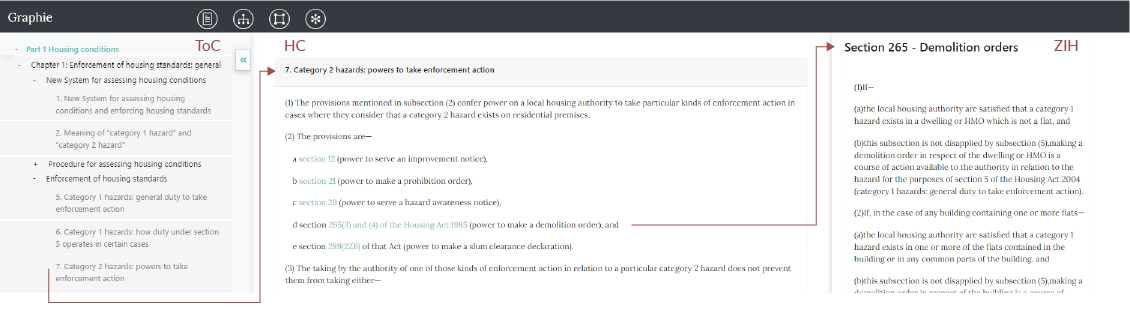}
	\caption{\label{fig:landingpage} An instance of Graphie's landing page. The user visualizes section 7 after clicking on the lowest link within the table of contents of Part 1, located on the left panel {\bf ToC}. Subsection 2(d) (within the middle panel, {\bf HC}) contains a hyperlinked outbound reference to section 265 of the Housing Act 1985. This hyperlink enables users to display section's 265 content on the right panel ({\bf ZIH}).}
\end{figure}

At this point of our data journey, we should expect that the original XML legislation files are now accurately represented on memory as an instance of the Graphie Python object, defined in section \ref{sec:graphiedatamodel}. In figure \ref{fig:landingpage}, we show on the left the table of contents generated using a specific instance of the Graphie object, mirroring the Housing Act's 2004 table of contents (footnote \ref{footnote:housing_act_2004_contents}). The JSON file feeding the graph visualization of figure \ref{fig:weight} is also generated based on the aforementioned Graphie object. 

The transformation service is a library of Python methods for serializing the memory representation of an Act into customized JSON or HTML outputs. Checking data integrity over this marshalling process is a crucial part of a data engineer's job. A data engineer will first automatically translate the on-memory data into a specified format, and then will consequently review the data validity of the obtained output. The relation between data integrity checks and the transformation service is also reflected in figure \ref{fig:pipeline}.

\begin{table}[ht]
\centering
    \begin{tabular}{ | l | l | p{5cm} |}
    \hline
    {\bf Element}  & {\bf Figures} & {\bf Method}  \\ \hline
    Table of contents ({\bf ToC}) &  Fig. \ref{fig:landingpage} &  \texttt{divNav()} \\ \hline
    Single Section ({\bf HC}) & Fig. \ref{fig:weight} , Fig. \ref{fig:landingpage} &  \texttt{htmlSingleSection()} \\ \hline
    Inbound Complexity Graph, JSON file & Fig. \ref{fig:weight} & \texttt{refInSingleLine()}, \texttt{refInSection()} \\ \hline
    Outbound Complexity Graph, JSON file & Fig. \ref{fig:networkgraphs} & \texttt{actsInSingleLine()}, \texttt{actsInSection()} \\ \hline
    \end{tabular}
    \caption{\label{table:transformation} Methods used to extract elements of the full data XML files, and where the results are visualized in this manuscript. In fig. \ref{fig:weight}, on the left, the table of contents is generated by the method \texttt{divNav()}. The method \texttt{htmlSingleSection()} instead prints one single section's content as a customized html paragraph of our platform. We use the methods \texttt{referencesInSection()} 
    and \texttt{actsInSection()} for picking out which nodes (sections) should be linked together with an edge in figures \ref{fig:weight} and \ref{fig:networkgraphs}.}
\end{table}

Let us take as an example the single line of subsection 2(a) in section 7 (figure \ref{fig:landingpage}): ``section 12 (power to serve an improvement notice)''. Use this line as an input parameter to \texttt{refInSingleLine()}, and \texttt{section 12} is returned. Thus, section 7 and section 12 form an {\bf inbound reference} that we also express as a dedicated link between node 7 (section 7) and node 12 (section 12) in figure \ref{fig:weight}. Passing section 7 to \texttt{refInSection()} returns the following array: {\tt [\textquote{section 12}:1, \textquote{section 21}:1, \textquote{section 29}:1]}. The array indicates three single {\bf inbound references}, from section 2 to sections 12, 21 and 29. Such calculated metadata values are also included in {\texttt{inbound.json}}\footnote{\url{graphie.quantlaw.co.uk/inbound.json}}, the underlying file of the inbound complexity network, in figure \ref{fig:weight}.

Subsection 4(b)'s text in section 194 (figure \ref{fig:194}) is now passed as a parameter to \texttt{actsInSingleLine()} and {\tt \textquote{the Housing and Regeneration Act 2008}} title is returned. Thus, section 194 contains an outbound reference to the Act just mentioned. Pass the whole content of section 194 as an input to \texttt{actsInSection()} and you will obtain the following array: {\tt [\textquote{Housing Act 1985}:4, \textquote{Housing and Regeneration Act 2008}:1]}, highlighting all the {\bf outbound references} within section 194. The value $4$ next to the key {\tt the Housing Act 1985} is a calculated metadata value that counts how many times the Housing Act 1985 is textually mentioned within section 194. The same value determines that edge's weight (henceforth, thickness) in figure \ref{fig:networkgraphs}. Each edge's thickness is also represented in {\texttt{outbound.json}} by the attribute \texttt{`thick'} (same attribute is used in {\texttt{inbound.json}}). Thus, the method \texttt{actsInSection()}, through {\texttt{outbound.json}}, populates information about nodes and links shown in figure \ref{fig:networkgraphs}, a network that highlights the outbound connections between sections in Housing Act 2004 with sections from other Acts.

UserMark \footnote{\url{http://www.austlii.edu.au/techlib/usermark/}} is a similar and well matured mark up tool, which allows users to create hypertext links from any text to AustLII's\footnote{\url{http://www.austlii.edu.au/}} legislation and High Court corpus.

\section{Visualization} \label{sec:graphie}

In this section, we illustrate how {\bf Graphie} could be used in practice. {\bf Graphie} offers a few main functionalities: a One Page View ({\bf OPV}) navigation, and two network visualizations. These representations could be a powerful tool for portraying the legal understanding of an ideal law practitioner that knows all the interconnections and shortcuts among different sections and Acts of the UK’s Statute Book. While similar visualizations could be obtained by using native Python or R network packages, like \textquote{networkx} \footnote{\url{https://networkx.github.io/}}, network analyses are usually published using browser friendly tools, such as D3.js, Cytoscape\footnote{\url{https://cytoscape.org}}, KeyLines \footnote{\url{https://cambridge-intelligence.com/keylines/}}. For example in \cite{bioTool}, initial network analyses were completed in \textquote{networkx} whereas their output was produced in Cytoscape. Other examples are networks in Sampo-UI, which exploit Cytoscape.js \footnote{\url{https://manual.cytoscape.org/en/stable/Navigation_and_Layout.html}} mechanisms and features: zooming, panning, node sizing, node coloring, and directed edges. In {\bf Graphie} we implemented, using a native D3 library, a novel network navigation framework that allows users to click on a node and display that node's underlying textual content. Such advanced mouse functionalities are not supported by default in Cytoscape.js and required further development.

\subsection{One Page View}

The landing page of {\bf Graphie}, depicted in Figure~\ref{fig:landingpage} offers a {\bf One-Page View} ({\bf OPV}) of the Housing Act 2004. {\bf OPV} displays a three-part canvas made up of the following panels: Table of Contents ({\bf ToC}), highlighted contents ({\bf HC}), and zoomed-in hyperlinks ({\bf ZIH}). On the left, users can explore the Housing Act's 2004 Table of Contents ({\bf ToC}) by expanding the related tree hierarchy. By clicking on a {\bf ToC} link, the user can see the actual content of the chosen section in the next panel ({\bf HC}). Let us go over an example: A user clicks on section's 7 {\bf ToC} link (figure \ref{fig:landingpage}) and that section's content is shown within the {\bf HC} area. Section 7 is connected with several other sections, and the user may now click on the link of section 265 of the Housing Act 1985 (outbound reference). This section's content is then pulled up on the right {\bf ZIH} panel. All mentioned panels ({\bf ToC}, {\bf HC}, {\bf ZIH}) are visible at all times. Thus, {\bf OPV} enables users to review different sections on one single page, without switching between different pages (as experienced in \cite{UKLegislation}\cite{Westlaw}). Also, sections are not hyperlinked in one Act's ``full text'' page on \url{legislation.gov.uk} (see footnote \ref{footnote:housing_act_2004}). Further, the {\bf OPV} feature is not offered in Westlaw \cite{Westlaw}. In {\bf Graphie} instead, section numbers are hyperlinked.

From a technical point view, {\bf ToC} displays the table of contents of the Housing Act 2004 as a nested ordered list. The {\bf HTML} content of {\bf ToC} is generated by the method  \texttt{divNav()}, whereas sections in {\bf HC} and {\bf ZIH} are built by the method \texttt{htmlSingleSection()}. All mentioned methods are part of the Transformation service, see section \ref{sec:transformation}.

\subsection{Network View - Inbound Complexity}

By clicking on the four-square navigation icon in figure~\ref{fig:weight}, a network graph representation of the {\bf Housing Act 2004} is opened. In this network, only sections of the same Act (inbound references) are connected and displayed. Each section's raw XML data is analyzed for cross-references using the method \texttt{refInSingleLine()}, defined in section~\ref{sec:transformation}. We use the method's output for creating the json file \texttt{inbound.json} that includes clear information about this network's nodes and their connections. The same file is used for populating our network front-end library with data. The obtained network consists of 233 nodes and 395 edges. In \texttt{inbound.json}, each node record and each link record are associated with metadata variables, such as  ``nodeSize'' and ``thick''.  Edge thickness in figure~\ref{fig:weight}, depends on ``thick'' values (also defined in section~\ref{sec:transformation}). Node size instead is set according to ``nodeSize'' and reflects the total count of references found within the Housing Act 2004 about that node's section number. That is, section 194 appears only once in section 270, subsection 5(c). Thus, section 194's node size is 1. 

Node coloring works as follows: each Part is assigned with a colour, and any nodes within this Part should be assigned the same color. For instance, section 1 of Part 1 is denoted as the blue node 1. All sections (nodes) of Part 1 are colored in blue. The user can  still click on one node's circle, and display their hyperlinked content on the right ({\bf ZIH} panel).

The network reports strong connectivity between same colored nodes, indicating that sections of the same {\bf Part} are expected to be tightly linked together. Sections of Part 5 are content-less, thus excluded from this inbound complexity network. 

\subsection{Network View - Outbound Complexity}

\begin{figure}
	\centering
	\includegraphics[width=1 \textwidth]{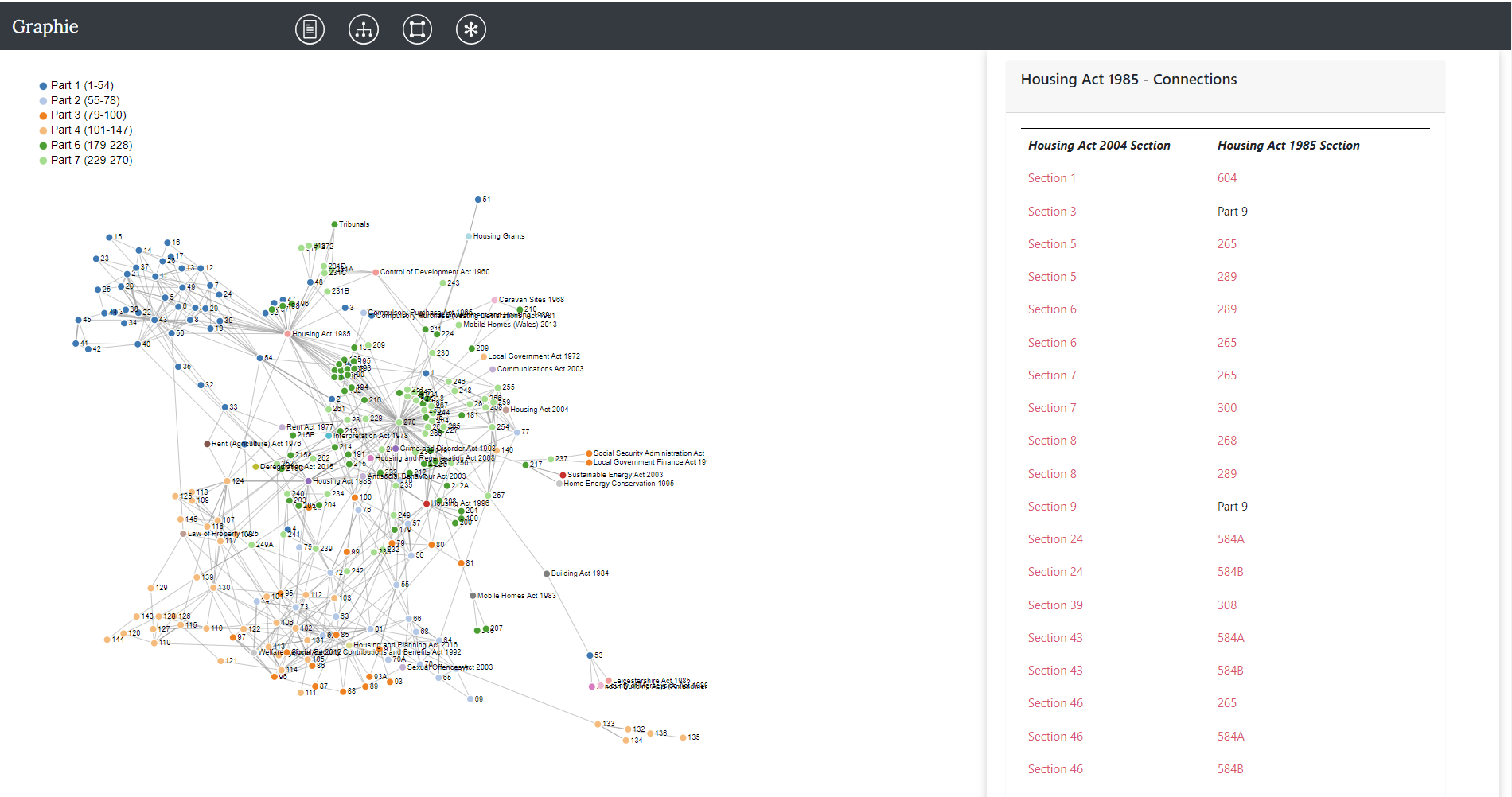}
	\caption{\label{fig:networkgraphs} Our full legal network representation of the Housing Act 2004 in which nodes are connected with others they share inbound or outbound references with. In this instance, the user could click on ``the Housing Act 1985'' node and fetch on the right ({\bf ZIH} panel) all the outbound references connecting the Housing Act 2004 with the aforementioned Act. Section numbers are hyperlinked and users can continue exploring sections from both Acts. }
\end{figure}

Table \ref{table:externalacts} lists the 5 most commonly mentioned Acts within the Housing Act's 2004 sections. There are 39 such Acts, obtained by applying the method \texttt{actsInSingleLine()} to each section of the Housing Act 2004. Each of these Acts is represented as a single, colored node in figure \ref{fig:networkgraphs}.
This new network is obtained from the original inbound network (figure \ref{fig:weight}) by also connecting sections of the Housing Act 2004 with the collected new (external) Acts. The resulting graph consists of 282 nodes and 673 edges and is obtained in Graphie by clicking on the star navigation icon. The user again can click on a node and display the underlying content of the chosen section. Network statistics and observables will be collected and analyzed in a forthcoming publication.

\begin{table}[ht]
\centering
    \begin{tabular}{ | l | l | p{5cm} |}
    \hline
    {\bf Act Name}  & {\bf \texttt{actsInSection()}} & {\bf Data Integrity}  \\ \hline
        Housing Act 1985 & 51 &  62 \\ \hline
        Housing Act 1988 & 18  &  20 \\ \hline
        Housing Act 1996 &  17 &  17 \\ \hline
        Housing and Planning Act 2016 &  15 &  24 \\ \hline
    \end{tabular}
    \caption{\label{table:externalacts} Table showing the top 4 most frequent external Acts connected with sections of the Housing Act 2004. For each Act, the column {\bf \texttt{actsInSection()}} displays those occurrences found by calling the previous method (see section \ref{sec:transformation}) whereas the next column ``Data Integrity'' is meant to show the actual observed occurrences after our data integrity checks (say, regular expression searches elaborated in section \ref{sec:dataintegirty} and table \ref{table:dataintegritytable})}
\end{table}

\section{Conclusions} \label{sec:conclusions}

In this paper, we designed and described a cross-Act pipeline for establishing a connection between the raw data provided by the UK Legislation API and our platform's front-end network representations of Acts and Bills included in the UK's Statute Book. Our networks reveal interesting associations between individual provisions, and also enable users to fully explore one Act's content by just hovering  over each node with the mouse. Nodes are associated with their underlying textual, hyperlinked content, and by clicking on one node the user can display this information on a dedicated panel alongside. Visiting different sections and hopping between provisions of an Act no longer requires opening new pages via hyperlinks (or via copy-pasting of the new node's address into the browser): the user can now remain on the same page and arrive at the sought information by simply clicking or hovering on the nodes of interest.

In our efforts to mold legislation from raw data into their final network visualizations, we faced the task to process them carefully and accurately. We tried to automate this process as much as possible. Due to the structural variety of our underlying files and the need to calculate network-specific metadata instances (not included in our original data), we had to incorporate a supervised layer of data integrity and data quality checks. 

The next challenge now will be to apply the developed pipeline on a larger scale, and go beyond the Housing Act we focused on for illustrative purposes. There are several ideas and further steps to be considered. First and foremost, we wish to use and test our parsing and data integrity routines against a larger volume and broader classes of Acts. The process of identifying references between sections (inbound or outbound) and populating related hyperlinks should be further improved and automated to reduce the need for human supervision. Due to the expected larger population of nodes, we might consider adjusting or creating new visualisations tools to avoid cramming effects on the screen. Such refinements usually require long learning and development curves. Our experience working with the sections of the Housing Act 2004 has however convincingly demonstrated the proof of concept of a fully operational prototype for the network-based visualization of the UK's Primary Legislation -- beyond the classical text-only paradigm -- which promises to offer a fresh way to conceive, analyze, and represent legal texts in a user-friendly way. 

\section{Data availability}

All data used was publicly available and downloaded from the UK's Legislation platform \cite{UKLegislation}. 

\section{Software availability} \label{sec:software}

Graphie is published at: \click{https://graphie.quantlaw.co.uk}{graphie.quantlaw.co.uk}. The code used in this study is available on Github, in the following repositories: 

\begin{itemize}
	\item Sofia, a Jupyter Pipeline: https://github.com/kclquantlaw/pipeline
	\item Web Platform's source code: https://github.com/kclquantlaw/graphie
\end{itemize}

\section{Acknowledgments}
PV and ET acknowledge support from UKRI Future Leaders Fellowship scheme [n.
MR/S03174X/1]. Y-PF is supported by the EPSRC Centre for Doctoral Training in Cross-disciplinary Approaches to Non-Equilibrium Systems (CANES EP/L015854/1).


\begin{thebibliography}{26}
\providecommand{\natexlab}[1]{#1}
\providecommand{\url}[1]{\texttt{#1}}
\expandafter\ifx\csname urlstyle\endcsname\relax
  \providecommand{\doi}[1]{doi: #1}\else
  \providecommand{\doi}{doi: \begingroup \urlstyle{rm}\Url}\fi

\bibitem[Loft and Apostolova(2017)]{LegislationVolume}
P.~Loft and V.~Apostolova.
\newblock {Acts and Statutory Instruments: the volume of UK legislation 1950 to
  2016}.
\newblock
  \url{https://researchbriefings.files.parliament.uk/documents/CBP-7438/CBP-7438.pdf},
  2017.
\newblock 

\bibitem[Archives(2022{\natexlab{a}})]{UKLegislation}
The National Archives.
\newblock {Enacted UK Legislation}.
\newblock \url{https://www.legislation.gov.uk/}, 2022{\natexlab{a}}.
\newblock 

\bibitem[AustLII(2022{\natexlab{b}})]{AustLII}
Australasian Legal Information Institute.
\newblock \url{http://www.austlii.edu.au/}, 2022{\natexlab{b}}.
\newblock 

\bibitem[Reuters(2022)]{Westlaw}
Thomson Reuters.
\newblock {Westlaw UK - Online Legal Research}.
\newblock \url{https://legal.thomsonreuters.com/en/westlaw}, 2022.
\newblock 

\bibitem[Ruhl and Katz(2015)]{LegalMaps}
J.B. Ruhl and D. M. Katz.
\newblock Measuring, monitoring, and managing legal complexity.
\newblock \emph{Iowa L. Rev.}, 101:\penalty0 191, 2015.

\bibitem[Koniaris et~al.(2017)Koniaris, Anagnostopoulos, and
  Vassiliou]{LegislationNetwork}
M. Koniaris, I. Anagnostopoulos, and Y. Vassiliou.
\newblock Network analysis in the legal domain: a complex model for European
  Union legal sources.
\newblock \emph{J Complex Netw.}, 6:243, 2017.

\bibitem[Vivo et~al.(2021)Vivo, Katz, and Ruhl]{PhysicsoftheLaw}
P. Vivo, D.~M. Katz, and J.B. Ruhl, editors.
\newblock \emph{The Physics of the Law: Legal Systems Through the Prism of
  Complexity Science}, volume~9 of \emph{Front. Phys.}, 2021. Frontiers
  in Physics.

\bibitem[Smolyak and Havlin(2022)]{EconophysicsReview}
A. Smolyak and S. Havlin.
\newblock {Three Decades in Econophysics—From Microscopic Modelling to
  Macroscopic Complexity and Back}.
\newblock \emph{Entropy}, 24:\penalty0 271, 2022.

\bibitem[Codling et~al.(2008)Codling, Plank, and Benhamou]{RandomWalksBiology}
E. Codling, M. Plank, and S. Benhamou.
\newblock {Random walks in biology}.
\newblock \emph{J. R. Soc. Interface}, 5:\penalty0 813,
  2008.

\bibitem[Katz and Bommarito(2014)]{hierarchical}
D.~M. Katz and M.~J. Bommarito.
\newblock Measuring the complexity of the law: the united states code.
\newblock \emph{Artif. Intell. Law}, 22:\penalty0 337, 2014.

\bibitem[F\"orster et~al.(2022)F\"orster, Annibale, Gamberi, Tzanis, and
  Vivo]{LegalTrees}
Y.-P. F\"orster, A. Annibale, L. Gamberi, E. Tzanis, and
  P. Vivo.
\newblock {Information retrieval and structural complexity of legal trees}.
\newblock \emph{J. Phys.: Complexity}, 3:\penalty0 035008, 2022.

\bibitem[Coupette et~al.(2021)Coupette, Beckedorf, Hartung, Bommarito, and
  Katz]{LawTime}
C.~Coupette, J.~Beckedorf, D.~Hartung, M.~Bommarito, and D.~M. Katz.
\newblock Measuring law over time: A network analytical framework with an
  application to statutes and regulations in the United States and Germany.
\newblock \emph{Front. Phys.} 9:658463, 2021.

\bibitem[Erdelez and O’Hare(1997)]{LegalInformatics}
S.~Erdelez and S.~O’Hare.
\newblock Legal informatics: Application of information technology in law.
\newblock \emph{Annu. Rev. Inf. Sci. Technol.}, 32:367, 1997.

\bibitem[Hyv\"{o}nen et~al.(2021)Hyv\"{o}nen, Tamper, Ikkala, Koho, Leal,
  Kesäniemi, Oksanen, Tuominen, and Hietanen]{LawSampo}
E.~Hyv\"{o}nen, M.~Tamper, E.~Ikkala, M.~Koho, R.~Leal, J.~Kesäniemi, A.~Oksanen,
  J.~Tuominen, and A.~Hietanen.
\newblock Lawsampo portal and data service for publishing and using legislation
  and case law as linked open data on the semantic web.
\newblock \emph{submitted}, 
\newblock \url{https://seco.cs.aalto.fi/publications/}, 2021.

\bibitem[Wilkinson et~al.(2016)Wilkinson, Dumontier, Aalbersberg, Appleton,
  Axton, Baak, Blomberg, Boiten, da~Silva~Santos, Bourne, and others.]{Fair}
M. Wilkinson, M. Dumontier, I. Aalbersberg \emph{et al}.
\newblock The {FAIR} guiding principles for scientific data management and stewardship.
\newblock \emph{Sci. Data} 3:160018, 2016.

\bibitem[Hyv\"{o}nen(2021)]{SampoModel}
E.~Hyv\"{o}nen.
\newblock Digital humanities on the semantic web: Sampo model and portal
  series.
\newblock \emph{Semant. Web, accepted}, 2022.

\bibitem[Ikkala et~al.(2021)Ikkala, Hyv\"{o}nen, Rantala, and Koho]{SampoUI}
E.~Ikkala, E.~Hyv\"{o}nen, H.~Rantala, and M.~Koho.
\newblock Sampo-ui, a full stack javascript framework for developing semantic
  portal user interfaces. semantic web – interoperability, usability,
  applicability.
\newblock \emph{Semant. Web} 13:69, 2022.

\bibitem[Robinson et~al.(2009)Robinson, Yu, Zeller, and Felten]{GovernmentData}
D.~Robinson, H.~Yu, W.~Zeller, and E.~Felten.
\newblock Government data and the invisible hand.
\newblock \emph{Yale J. L. Tech.}, 11:\penalty0 159, 2009.

\bibitem[A. et~al.(2019)A., J., E., M., Hietanen, and
  A.~Hyv\"{o}nen]{SemanticFinlex}
A. Oksanen, J. Tuominen, E. M\"{a}kel\"{a}, M. Tamper, A. Hietanen, and
  E.~Hyv\"{o}nen.
\newblock Semantic finlex: Transforming, publishing, and using finnish
  legislation and case law as linked open data on the web.
\newblock \emph{In: Peruginelli, G., Faro, S. (eds.) Knowledge of the Law in
  the Big Data Age, Front. Artif. Intell. Appl.},
  317:\penalty0 212, 2019.

\bibitem[Lewis~Mcgibbney(2013)]{CLML}
B.~Kumar L.~Mcgibbney.
\newblock {A comparative study to determine a suitable representational data
  model for UK building regulations}.
\newblock \emph{J. Inf. Tech. Construction}, 18:\penalty0 20, 2013.

\bibitem[Lab(2022)]{tamingthecomplexity}
King's~Digital Lab.
\newblock {Taming the complexity of law | Modelling and visualization of
  dynamically interacting legal systems}.
\newblock \url{https://kingsdigitallab.github.io/tcl/}, 2022.
\newblock 

\bibitem[Archives(2022{\natexlab{c}})]{HousingActXML}
The National Archives.
\newblock {Housing Act 2004, Section 194, Enacted UK Legislation}.
\newblock \url{https://www.legislation.gov.uk/ukpga/2004/34/data.xml},
  2022{\natexlab{c}}.
\newblock 

\bibitem[Archives(2022{\natexlab{d}})]{LegislationStructure}
The National Archives.
\newblock {FAQ, Enacted UK Legislation}.
\newblock \url{https://www.legislation.gov.uk/help}, 2022{\natexlab{d}}.
\newblock 

\bibitem[Parliamentary Counsel’s~Office(2022)]{ReadingLegislation}
Government of Western Australia, Department of the Attorney~General,
  Parliamentary Counsel’s~Office.
\newblock {How to read legislation, a beginner's guide}.
\newblock
  \url{https://www.legislation.wa.gov.au/legislation/statutes.nsf/RedirectURL?OpenAgent&query=howtoreadlegislation.pdf},
  2022.
\newblock 

\bibitem[Ferolito et~al.(2014)Ferolito, do~Valle, Gerlovin, Costa, Casas,
  Gaziano, et~al.]{bioTool}
B.~Ferolito, I. F.~do~Valle, H.~Gerlovin \emph{et al.}
\newblock Visualizing novel connections and genetic similarities across
  diseases using a network‑medicine based approach.
\newblock \emph{Scientific Reports}, 12:14914, 2022.

\end{thebibliography}
\end{document}